\documentclass{aa}  
\usepackage {graphicx}
\usepackage {times}
\usepackage {natbib}

\begin{document}

   \title{Evolution of Infrared Luminosity functions of Galaxies in the AKARI NEP-Deep field}

   \subtitle{Revealing the cosmic star formation history hidden by dust\thanks{This research is based on the observations with AKARI, a JAXA project with the participation of ESA.},\thanks{Based  on data collected at Subaru Telescope, which is operated by the National Astronomical Observatory of Japan.}}
\titlerunning{Infrared Luminosity functions with the AKARI}

   \author{Tomotsugu Goto
          \inst{1,2}\fnmsep\thanks{JSPS SPD fellow},
          T.Takagi\inst{3},
	  H.Matsuhara\inst{3},
	  T.T.Takeuchi\inst{4},
	  C.Pearson\inst{5,6,7},
	  T.Wada \inst{3},
	  T.Nakagawa\inst{3},
	  O.Ilbert\inst{8},
	  E.Le Floc'h\inst{9},
	  S.Oyabu\inst{3},
	  Y.Ohyama\inst{10},
	  M.Malkan\inst{11},
          H.M.Lee\inst{12},   
          M.G.Lee\inst{12},
	  H.Inami\inst{3,13,14},
          N.Hwang\inst{2},
H.Hanami\inst{15},
M.Im\inst{12},
K.Imai\inst{16},
T.Ishigaki\inst{17},
S.Serjeant\inst{7},
\and
H. Shim \inst{12}
          }

	  \institute{
	  Institute for Astronomy, University of Hawaii,	  2680 Woodlawn Drive, Honolulu, HI, 96822, USA\\
	       \email{tomo@ifa.hawaii.edu}
	       \and
	  National Astronomical Observatory, 2-21-1 Osawa, Mitaka, Tokyo, 181-8588,Japan
		      \and
          Institute of Space and Astronautical Science, Japan Aerospace Exploration Agency, 	     Sagamihara, Kanagawa 229-8510
\and Institute for Advanced Research, Nagoya University, Furo-cho, Chikusa-ku, Nagoya 464-8601
\and Rutherford Appleton Laboratory, Chilton, Didcot, Oxfordshire OX11 0QX, UK
\and Department of Physics, University of Lethbridge, 4401 University Drive,Lethbridge, Alberta T1J 1B1, Canada
\and Astrophysics Group, Department of Physics,  The Open University, Milton Keynes, MK7 6AA, UK
\and Laboratoire d'Astrophysique de Marseille, BP 8, Traverse du Siphon, 13376 Marseille Cedex 12, France
\and CEA-Saclay, Service d'Astrophysique, France
\and Academia Sinica, Institute of Astronomy and Astrophysics, Taiwan
\and Department of Physics and Astronomy, UCLA, Los Angeles, CA, 90095-1547 USA
\and Department of Physics \& Astronomy, FPRD, Seoul National University, Shillim-Dong, Kwanak-Gu, Seoul 151-742, Korea	
\and Spitzer Science Center, California Institute of Technology, Pasadena, CA 91125
\and Department of Astronomical Science,The Graduate University for Advanced Studies
\and Physics Section, Faculty of Humanities and Social Sciences, Iwate University, Morioka, 020-8550
\and  TOME R\&D Inc. Kawasaki, Kanagawa 213 0012, Japan
\and Asahikawa National College of Technology, 2-1-6 2-jo Shunkohdai, Asahikawa-shi, Hokkaido 071-8142 
	     }
   \date{Received September 15, 2009; accepted December 16, 2009}
   \authorrunning{Goto et al.}

 
  \abstract
{}
   {Dust-obscured star-formation becomes much more important with increasing intensity, and increasing redshift. 
 We aim to reveal cosmic star-formation history obscured by dust using deep infrared observation with the AKARI.
   }
   {We construct restframe 8$\mu$m,  12$\mu$m, and total infrared (TIR) luminosity functions (LFs) at $0.15<z<2.2$ using  4128 infrared sources  in the AKARI NEP-Deep field.
A continuous filter coverage in the mid-IR wavelength (2.4, 3.2, 4.1, 7, 9, 11, 15, 18, and 24$\mu$m) by the AKARI satellite allows us to estimate restframe 8$\mu$m and 12$\mu$m luminosities without using a large extrapolation based on a SED fit, which was the largest uncertainty in previous work. 
}
   {
We have found that all 8$\mu$m ($0.38<z<2.2$), 12$\mu$m ($0.15<z<1.16$), and TIR LFs ($0.2<z<1.6$), show a continuous and strong evolution toward higher redshift. 
 In terms of cosmic infrared luminosity density ($\Omega_{IR}$), which was obtained by integrating analytic fits to the LFs, we found a good agreement with previous work at $z<1.2$. 
 We found the $\Omega_{IR}$ evolves as $\propto (1+z)^{4.4\pm 1.0}$.
 When we separate contributions to $\Omega_{IR}$ by LIRGs and ULIRGs, we found more IR luminous sources are increasingly more important at higher redshift.  We found that the ULIRG (LIRG) contribution increases by a factor of 10 (1.8) from $z$=0.35 to $z$=1.4.
}
   {}


  \maketitle

\section{Introduction}

Studies of the extragalactic background suggest at least half the luminous energy generated by stars has been reprocessed into the infrared (IR) by dust \citep{1999A&A...344..322L,1996A&A...308L...5P,2008A&A...487..837F}, suggesting that dust-obscured star formation was much more important at higher redshifts than today.

\citet{2005ApJ...625...23B} estimate that IR luminosity density is 7 times higher than the UV luminosity density at z$\sim$0.7 than locally.
 \citet{2005A&A...440L..17T} reported that UV-to-IR luminosity density ratio, $\rho_{L(UV)}/\rho_{L(dust)}$, 
 evolves from 3.75 ($z$=0) to 15.1 by $z$=1.0 with a careful treatment of the sample selection effect, and that 70\% of star formation activity is obscured by dust at 0.5$<z<$1.2.
Both works highlight the importance of probing cosmic star formation activity at high redshift in the infrared bands. 
Several works found that most extreme star-forming (SF) galaxies, which are increasingly important at higher redshifts, are also more heavily obscured by dust \citep{2001AJ....122..288H,2001ApJ...558...72S,2007ApJS..173..404B}.

Despite the value of infrared observations,
 studies of infrared galaxies by the IRAS and the ISO were restricted to bright sources due to the limited sensitivities \citep{1990MNRAS.242..318S,1997MNRAS.289..490R,1999ApJ...517..148F,2004MNRAS.355..813S,2006A&A...448..525T,2003ApJ...587L..89T}, until the recent launch of the Spitzer and the AKARI satellites.
 Their enormous improved sensitivities have revolutionized the field. For example:
%

\citet{2005ApJ...632..169L} analyzed the evolution of the total and 15$\mu$m IR luminosity functions (LFs) at $0<z<1$ based on the the Spitzer MIPS 24$\mu$m data ($>83\mu$Jy and $R<24$)  in the CDF-S, and found a positive evolution in both luminosity and density, suggesting increasing importance of the LIRG and ULIRG populations at higher redshifts. 

\citet{2005ApJ...630...82P} used MIPS 24$\mu$m observations of the CDF-S and HDF-N ($>83\mu$Jy) to find that
 that $L^*$ steadily increases by an order of magnitude to $z\sim 2$, suggesting that the luminosity evolution is stronger than the density evolution. The $\Omega_{TIR}$ scales as (1+z)$^{4.0\pm 0.2}$ from $z$=0 to 0.8.

\citet{2006MNRAS.370.1159B} constructed LFs at 3.6, 4.5, 5.8, 8 and 24$\mu$m over $0<z<2$ using the data from the Spitzer Wide-area Infrared Extragalactic (SWIRE) Survey in a 6.5 deg$^2$ ($S_{24 \mu m}>230 \mu$Jy). 
They found a clear luminosity evolution in all the bands, but the evolution is more pronounced at longer wavelength; extrapolating from
24$\mu$m, they inferred that  $\Omega_{TIR}\propto$(1+z)$^{4.5}$.
They constructed separate LFs for three different galaxy SED (spectral energy distribution) types and Type 1 AGN, finding that starburst and late-type galaxies showed stronger evolution.
Comparison of 3.6 and 4.5$\mu$m LFs with semi-analytic and spectrophotometric models suggested that the IMF is skewed towards higher mass star formation in 
more intense starbursts.

\citet{2007ApJ...660...97C} estimated restframe 8$\mu$m LFs  of galaxies over 0.08deg$^2$ in the GOODS fields based on Spitzer 24$\mu$m ($>80\mu$Jy) at $z$=1 and 2.
They found  a continuous and strong positive luminosity evolution from $z$=0 to $z$=1, and to $z$=2. 
However, they also found that the number density of star-forming galaxies with $\nu L_{\nu}^{8\mu m}>10^{10.5} L_{\odot}$ (AGNs are excluded.) increases by a factor of 20 from $z$=0 to 1, but decreases by half from $z$=1 to 2 mainly due to the decrease of LIRGs.

\citet{2009A&A...496...57M} investigated restframe 15$\mu$m, 35$\mu$m and total infrared (TIR) LFs using deep 70$\mu$m observations ($\sim$300 $\mu$Jy) in the Spitzer GOODS and FIDEL (Far Infrared Deep Extragalactic Legacy Survey) fields (0.22 deg$^2$ in total) at $z<1.3$. 
 They stacked 70$\mu$m flux at the positions of 24$\mu$m sources when sources are not detected in 70$\mu$m.
 They found no change in the shape of the LFs, but found a pure luminosity evolution proportional to (1+z)$^{3.6\pm 0.5}$, and that  LIRGs and ULIRGs have increased by a factor of 40 and 100 in number density by $z\sim$1. 

Also, see \citet{2009ApJ...697..506D} for 3.6-8.0 $\mu$m LFs based on the IRAC photometry in the NOAO Deep Wide-Field Survey Bootes field.

However, most of the Spitzer work relied on a large extrapolation from 24$\mu$m flux to estimate the 8, 12$\mu$m or TIR luminosity.
Consequently,  Spitzer results heavily depended on the assumed IR SED library \citep{2002ApJ...576..159D,2003MNRAS.338..555L,2001ApJ...556..562C}. 
Indeed many authors pointed out that the largest uncertainty in these previous IR LFs came from SED models,  especially when one computes TIR luminosity solely from observed 24$\mu$m flux  \citep[e.g., see Fig. 5 of][]{2007ApJ...660...97C}.

AKARI, the first Japanese IR dedicated satellite, has continuous filter coverage across the mid-IR wavelengths,  thus, allows us to estimate MIR (mid-infrared)-luminosity without using a large $k$-correction based on the SED models, eliminating the largest uncertainty in previous work. 
By taking advantage of this, we present the restframe 8, 12$\mu$m and TIR LFs using the AKARI NEP-Deep data in this work.
 
Restframe 8$\mu$m luminosity in particular is of primary relevance for star-forming galaxies, as it includes polycyclic aromatic hydrocarbon (PAH) emission. PAH molecules characterize star-forming regions \citep{1990A&A...237..215D}, and the associated emission lines between 3.3 and 17 $\mu $m dominate the SED of star-forming galaxies  with a main bump located around 7.7$\mu$m. Restframe 8$\mu$m luminosities have been confirmed to be good indicators of knots of star formation \citep{2005ApJ...633..871C}
and of the overall star formation activity of star forming galaxies \citep{2005ApJ...632L..79W}.
 At $z$=0.375, 0.875, 1.25 and 2, the restframe 8$\mu$m is covered by the AKARI 
 $S11$, $L15,L18W$ and $L24$ filters. We present  the restframe 8$\mu$m LFs at these redshifts at Section\ref{sec:8umlf}.

Restframe 12$\mu$m luminosity functions have also been studied extensively \citep{1993ApJS...89....1R,2005ApJ...630...82P}.
 At $z$=0.25, 0.5 and 1, the restframe 12$\mu$m is covered by the AKARI  $L15,L18W$ and $L24$ 
 filters. We present  the restframe 12$\mu$m LFs at these redshifts in Section\ref{sec:12umlf}.

We also estimate TIR LFs through the SED fit using all the mid-IR bands of the AKARI. The results are presented in Section \ref{sec:tirlf}.

  Unless otherwise stated, we adopt a cosmology with $(h,\Omega_m,\Omega_\Lambda) = (0.7,0.3,0.7)$ \citep{2008arXiv0803.0547K}.

\section{Data \& Analysis}
\label{Data}


\begin{figure}
\begin{center}
\includegraphics[scale=0.3]{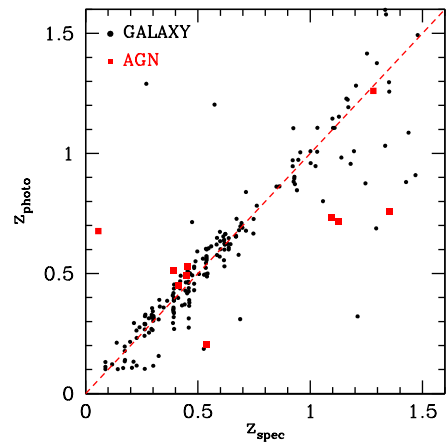}
\end{center}
\caption{
Photometric redshift estimates with {\ttfamily  LePhare} \citep{2006A&A...457..841I,2007A&A...476..137A,2009ApJ...690.1236I} for spectroscopically observed galaxies with Keck/DEIMOS (Takagi et al. in prep.).
Red squares show objects where AGN templates were better fit.
Errors of the photoz is $\frac{\Delta z}{1+z}$=0.036 for $z\leq 0.8$, but becomes worse at $z> 0.8$ to be  $\frac{\Delta z}{1+z}$=0.10 due mainly to the relatively shallow near-IR data.
}\label{fig:photoz}
\end{figure}

\begin{figure}
\begin{center}
\includegraphics[scale=0.6]{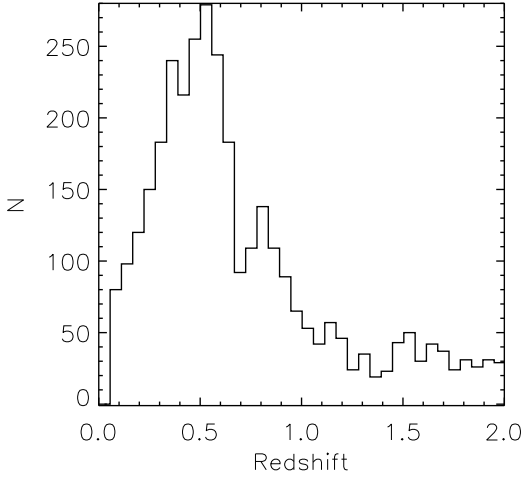}
\end{center}
\caption{
Photometric redshift distribution.
}\label{fig:photoz_hist}
\end{figure}

\begin{figure}
\begin{center}
\includegraphics[scale=0.6]{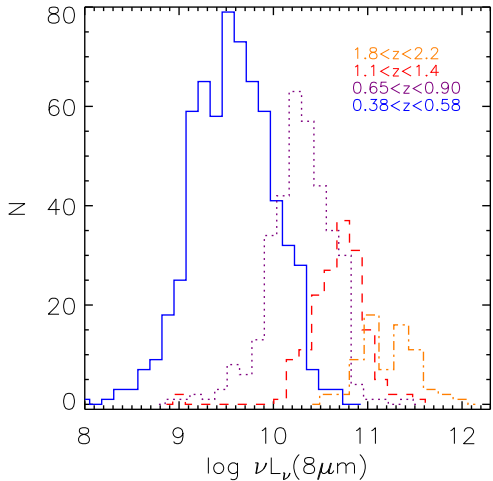}
\end{center}
\caption{
8$\mu$m luminosity distributions of samples used to compute  restframe  8$\mu$m LFs. 
From low redshift, 533, 466, 236 and 59 galaxies are in each redshift bin.
}\label{fig:8um_luminosity}
\end{figure}

\subsection{Multi-wavelength data in the AKARI NEP Deep field}
AKARI, the Japanese infrared satellite \citep{2007PASJ...59S.369M}, performed deep imaging in the North Ecliptic Region (NEP) from 2-24$\mu$m, with 14 pointings in each field over 0.4 deg$^2$ \citep{2006PASJ...58..673M,2007PASJ...59S.543M,2008PASJ...60S.517W}. Due to the solar synchronous orbit of the AKARI, the NEP is the only AKARI field with very deep imaging at these wavelengths. 
 The 5 $\sigma$ sensitivity in the AKARI IR filters ($N2,N3,N4,S7,S9W,S11,L15,L18W$ and $L24$) are 14.2, 11.0, 8.0, 48, 58, 71, 117, 121 and 275$\mu$Jy \citep{2008PASJ...60S.517W}. 
 These filters provide us with a unique continuous wavelength coverage at 2-24$\mu$m, 
where there is a gap between the Spitzer IRAC and MIPS, and the ISO $LW2$ and $LW3$.
Please consult \citet{2007PASJ...59S.515W,2008PASJ...60S.517W,Pearson18um,Pearson15um} for data verification and completeness estimate at these fluxes.
The PSF sizes are 4.4, 5.1, and 5.4'' in $2-4$, $7-11$, $15-24\mu$m bands. The depths of near-IR bands are limited by source confusion, but those of mid-IR bands are by sky noise.


In analyzing these observations, we first
combined the three images of the MIR channels, i.e. MIR-S($S7,S9W$, and $S11$) and 
MIR-L($L15,L18W$ and $L24$), in order to obtain  two high-quality images. In the 
resulting MIR-S and MIR-L images, the residual sky has been 
reduced significantly, which helps to obtain more reliable source catalogues. 
For both the MIR-S and MIR-L channels, we use SExtractor for the combined images to 
determine initial source positions. 

We follow \citet{2007PASJ...59S.557T} procedures for photometry and band-merging of IRC sources. 
But this time, to maximize the number of MIR sources,
we made two IRC band-merged catalogues based on the combined MIR-S and 
MIR-L images, and then concatenated these catalogues, eliminating duplicates. 

In the band-merging process, the source centroid in each IRC image has been 
determined, starting from the source position in the combined images as 
the initial guess. If the centroid determined in this way is shifted from the 
original position by $>3''$, we reject such a source as the counterpart. 
We note that this band-merging method is used only for IRC bands.

We compared raw number counts with previous work based on the same data but with different source extraction methods \citep{2008PASJ...60S.517W,Pearson18um,Pearson15um} and found a good agreement.

 A subregion of the NEP-Deep field was observed in the $BVRi'z'$-bands with the Subaru telescope \citep{2007AJ....133.2418I,2008PASJ...60S.517W}, reaching limiting magnitudes of $z_{AB}=$26  in one field of view of the Suprime-Cam. 
We restrict our analysis to the data in this Suprime-Cam field (0.25 deg$^2$), where we have enough UV-opical-NIR coverage to estimate good photometric redshifts.
 The $u'$-band photometry in this area is provided by the CFHT (Serjeant et al. in prep.).
 The same field was also observed with the KPNO2m/FLAMINGOs in $J$ and $Ks$ to the depth of $Ks_{Vega}<20$ \citep{2007AJ....133.2418I}. 
 GALEX covered the entire field to depths of $FUV<25$ and $NUV<25$ (Malkan et al. in prep.).

 In the Suprime-Cam field of the AKARI NEP-Deep field, there are a total of 4128 infrared sources down to $\sim$100 $\mu$Jy in the $L18W$ filter.
 All magnitudes are given in AB system in this paper.

For the optical identification of MIR sources, we adopt the likelihood
ratio (LR) method  \citep{1992MNRAS.259..413S}.
For the probability distribution 
functions of magnitude and angular separation based on correct optical 
counterparts (and for this purpose only), we use a subset of IRC sources, which are detected in all 
IRC bands. For this subset of 1100 all-band--detected sources, the optical 
counterparts are all visually inspected and ambiguous cases are excluded. 
There are multiple optical counterparts for 35\% of MIR sources within $<3''$. 
If we adopted the nearest neighbor approach for the optical identification, 
the optical counterparts differs from that of the LR method for 20\% of the
sources with multiple optical counterparts. Thus, in total we estimate 
that less than 15\% of MIR sources suffer from serious problems of optical 
identification.


\subsection{Photometric redshift estimation} \label{sec:photoz}

For these infrared sources, we have computed photometric redshift using a publicly available code, {\ttfamily LePhare }\footnote{http://www.cfht.hawaii.edu/$^{\sim}$arnouts/lephare.html} \citep{2006A&A...457..841I,2007A&A...476..137A,2009ApJ...690.1236I}.
The input magnitudes are  $FUV,NUV$(GALEX),$u$(CFHT), $B,V,R,i',z'$(Subaru), $J$, and $K$(KPNO2m).
 We summarize the filters used in Table \ref{tab:filters_used}.

Among various templates and fitting parameters we tried, we found the best results were obtained with the following: we used modified CWW \citep{1980ApJS...43..393C} and QSO templates. 
 These CWW templates are interpolated and adjusted to better match VVDS spectra \citep{2007A&A...476..137A}.
We included strong emission lines in computing colors. We used the Calzetti extinction law. 
 More details in training {\ttfamily LePhare} is given in \citet{2006A&A...457..841I}.
 
The resulting photometric redshift estimates agree reasonably well with 293 galaxies ($R<24$) with spectroscopic redshifts taken with Keck/DEIMOS in the NEP field (Takagi et al. in prep.).  
  The measured errors on the photo-$z$ are $\frac{\Delta z}{1+z}$=0.036 for $z\leq0.8$ and  $\frac{\Delta z}{1+z}$=0.10 for $z>0.8$.
 The $\frac{\Delta z}{1+z}$ becomes significantly larger at $z>0.8$, where we suffer from relative shallowness of our near-IR data.
 The rate of catastrophic failures is 4\% ($\frac{\Delta z}{1+z}>$0.2) among the spectroscopic sample.
 
 In Fig.\ref{fig:photoz}, we compare spectroscopic redshifts from Keck/DEIMOS (Takagi et al.) and our photometric redshift estimation. 
 Those  SEDs which are better fit with a QSO template are shown as red triangles. 
 We remove those red triangle objects  ($\sim$2\% of the sample) from the LFs presented below. 
 We caution that this can only remove extreme type-1 AGNs, and thus, fainter, type-2 AGN that could be removed by X-rays or optical spectroscopy still remain in the sample.  

 Fig.\ref{fig:photoz_hist} shows the distribution of photometric redshift. 
 The distribution has several peaks, which corresponds to galaxy clusters in the field \citep{2008PASJ...60S.531G}.
 We have 12\% of sources that do not have a good SED fit to obtain a reliable photometric redshift estimation.
 We apply this photo-$z$ completeness correction to the LFs we obtain.
 Readers are referred to \citet{2009MNRAS.394..375N}, who estimated photometric redshifts using only the AKARI filters to obtain 10\% accuracy.

\begin{table}
 \centering
 \begin{minipage}{180mm}
  \caption{Summary of filters used.}\label{tab:filters_used}
  \begin{tabular}{@{}ccclllcccc@{}}
  \hline
 Estimate &   Redshift & Filter \\ 
 \hline
 \hline
Photo $z$ &0.15$<$z$<$2.2 & $FUV,NUV$,$u$, $B,V,R,i',z$, $J$, and $K$\\
 \hline
 8$\mu$m LF &0.38$<$z$<$0.58  &    S11 (11$\mu$m) 	\\
 8$\mu$m LF &0.65$<$z$<$0.90  &    L15 (15$\mu$m) \\
 8$\mu$m LF &1.1$<$z$<$1.4    &    L18W (18$\mu$m) \\
 8$\mu$m LF &1.8$<$z$<$2.2    &    L24 (24$\mu$m)  \\
 \hline
 12$\mu$m LF &0.15$<$z$<$0.35  &    L15 (15$\mu$m) \\
 12$\mu$m LF &0.38$<$z$<$0.62  &     L18W (18$\mu$m) \\
 12$\mu$m LF &0.84$<$z$<$1.16  &    L24 (24$\mu$m)  \\
 \hline
TIR LF & 0.2$<$z$<$0.5 &  $S7,S9W,S11,L15,L18W$ and $L24$  \\
TIR LF &0.5$<$z$<$0.8  &  $S7,S9W,S11,L15,L18W$ and $L24$  \\
TIR LF &0.8$<$z$<$1.2  &  $S7,S9W,S11,L15,L18W$ and $L24$  \\
TIR LF &1.2$<$z$<$1.6  &  $S7,S9W,S11,L15,L18W$ and $L24$  \\
 \hline
\end{tabular}
\end{minipage}
\end{table}

\subsection{The 1/$V_{\max}$ method}\label{sec:vmax}

We compute LFs using the 1/$V_{\max}$ method \citep{1968ApJ...151..393S}. The advantage of the 1/$V_{\max}$  method is that it allows us to compute a LF directly from data, with no parameter dependence or an assumed model. A drawback is that it assumes a homogeneous galaxy distribution, and is thus vulnerable to local over-/under-densities \citep{2000ApJS..129....1T}. 

A comoving volume associated with any source of a given luminosity is defined as $V_{\max}=V_{z_{\max}}-V_{z_{\min}}$, where $z_{\min}$ is the lower limit of the redshift bin and $z_{\max}$ is the maximum redshift at which the object could be seen given the flux limit of the survey, with a maximum value corresponding to the upper redshift of the redshift bin. 
 More precisely,  
\begin{eqnarray}
 \mathrm{z_{\max} = min(z_{\max}~of~the~bin,~ z_{\max}~ from~ the ~flux~limit)}
\end{eqnarray}
We used the SED templates \citep{2003MNRAS.338..555L} for $k$-corrections to obtain the maximum observable redshift from the flux limit.

 For each luminosity bin then, the LF is derived as

\begin{eqnarray}
\phi =\frac{1}{\Delta L}\sum_{i} \frac{1}{V_{\max,i}}w_i, \label{LF}
\end{eqnarray}

\noindent where $V_{\max}$  is a comoving volume over which the $i$th galaxy could be observed, $\Delta L$ is the size of the luminosity bin (0.2 dex), and $w_i$ is the completeness correction factor of the $i$th galaxy. We use completeness correction measured by \citet{2008PASJ...60S.517W} for 11 and 24 $\mu$m and \citet{Pearson18um,Pearson15um} for 15 and 18$\mu$m. This correction is 25\% at maximum, since we only use the sample where the completeness is greater than 80\%.

\subsection{Monte Carlo simulation}\label{sec:monte}
 Uncertainties of the LF values stem from various factors such as fluctuations in 
 the number of sources in each luminosity bin, 
 the photometric redshift uncertainties,
 the $k$-correction uncertainties,
 and the flux errors. 
 To compute these errors we performed Monte Carlo simulations by creating 1000 simulated catalogs, where 
 each catalog contains the same number of sources, but we assign each source a new redshift following a Gaussian distribution centered at the photometric redshift with the measured dispersion of $\Delta z/(1+z)=$0.036 for $z\leq0.8$ and $\Delta z/(1+z)=$0.10 for $z>0.8$ (Fig.\ref{fig:photoz}).
 The flux of each source is also allowed to vary according to the measured flux error following a Gaussian distribution.
 For 8$\mu$m and 12$\mu$m LFs, 
 we can ignore the errors  due to the $k$-correction 
 thanks to the AKARI MIR filter coverage. 
 For TIR LFs, we have added 0.05 dex of error for uncertainty in the SED fitting following the discussion in \citet[][]{2009A&A...496...57M}.
 We did not consider the uncertainty on the cosmic variance here since the AKARI NEP field covers a large volume and has comparable number counts to other general fields \citep{2007AJ....133.2418I,2008ApJ...683...45I}. Each redshift bin we use covers $\sim 10^6$ Mpc$^3$ of volume.
 See \citet{2006PASJ...58..673M} for more discussion on the cosmic variance in the NEP field.
 These estimated errors are added to the Poisson errors in each LF bin in quadrature.

%
%

\section{Results}\label{results}

\subsection{8$\mu$m LF}\label{sec:8umlf}

Monochromatic 8$\mu$m luminosity ($L_{8\mu m}$) is known to correlate well with the TIR luminosity \citep{2006MNRAS.370.1159B,2007ApJ...664..840H}, especially for star-forming galaxies because the rest-frame 8$\mu$m flux are dominated by prominent PAH features such as at 6.2, 7.7 and 8.6 $\mu$m.
 
 Since the AKARI has continuous coverage in the mid-IR wavelength range, the restframe 8$\mu$m luminosity can be obtained without a large uncertainty in $k$-correction at a corresponding redshift and filter. For example, at $z$=0.375, restframe 8$\mu$m is redshifted into $S11$ filter.
 Similarly, $L15,L18W$ and $L24$ cover restframe 8$\mu$m at $z$=0.875, 1.25 and 2.
 This continuous filter coverage is an advantage to AKARI data. Often SED models are used to extrapolate from Spitzer 24$\mu$m flux in previous work, producing a source of the largest uncertainty. 
  We summarise filters used in Table \ref{tab:filters_used}.
 
 To obtain restframe 8$\mu$m LF, we applied a flux limit of F(S11)$<$70.9, F(L15)$<$117, F(L18W)$<$121.4, and F(L24)$<$275.8 $\mu$Jy at $z$=0.38-0.58, $z$=0.65-0.90, $z$=1.1-1.4 and $z$=1.8-2.2, respectively. 
These are the 5$\sigma$ limits measured in \citet{2008PASJ...60S.517W}. We exclude those galaxies whose SEDs are better fit with QSO templates ($\S$\ref{Data}).

 We use the completeness curve presented in \citet{2008PASJ...60S.517W}  and \citet{Pearson18um,Pearson15um}  to correct for the incompleteness of the detection. However, this correction is 25\% at maximum since the sample is 80\% complete at the 5$\sigma$ limit. Our main conclusions are not affected by this incompleteness correction.
 To compensate for the increasing uncertainty in increasing $z$, we use redshift binsize of 0.38$<z<$0.58, 0.65$<z<$0.90, 1.1$<z<$1.4,  and 1.8$<z<$2.2.  
 We show the $L_{8\mu m}$ distribution in each redshift range in Fig.\ref{fig:8um_luminosity}.
 Within each redshift bin, we use 1/$V_{\max}$ method to compensate for the flux limit in each filter.

\begin{figure}
\includegraphics[scale=0.6]{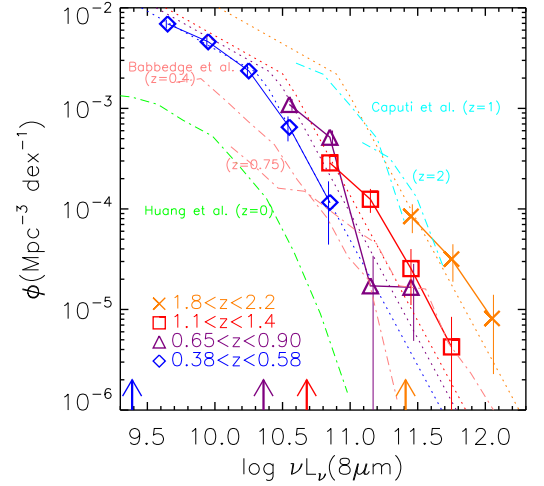}
\caption{
 Restframe  8$\mu$m LFs based on the AKARI NEP-Deep field.
 The blue diamons, purple triangles, red squares, and orange crosses show the 8$\mu$m LFs at $0.38<z<0.58, 0.65<z<0.90, 1.1<z<1.4$, and $1.8<z<2.2$, respectively. AKARI's MIR filters can observe restframe 8$\mu$m at these redshifts in a corresponding filter. Errorbars are from the Monte Caro simulations ($\S$\ref{sec:monte}).
 The dotted lines show analytical fits with a double-power law.
 Vertical arrows show the 8$\mu$m luminosity corresponding to the flux limit at the central redshift in each redshift bin.
 Overplotted are  \citet{2006MNRAS.370.1159B} in the pink dash-dotted lines, \citet{2007ApJ...660...97C} in the cyan dash-dotted lines, and \citet{2007ApJ...664..840H} in the green dash-dotted lines.
AGNs are excluded from the sample ($\S$\ref{sec:photoz}).
}\label{fig:8umlf}
\end{figure}

 We show the computed restframe 8$\mu$m LF in Fig.\ref{fig:8umlf}. 
 Arrows show the 8$\mu$m luminosity corresponding to the flux limit at the central redshift in each redshift bin.
 Errorbars on each point are based on the Monte Carlo simulation ($\S$ \ref{sec:vmax}).
 
 For a comparison, as the green dot-dashed line, we also show the 8$\mu$m LF of star-forming galaxies at $0<z<0.3$ by \citet{2007ApJ...664..840H}, using the 1/$V_{\max}$  method applied to the IRAC 8$\mu$m GTO data. 
 Compared to the local LF, our  8$\mu$m LFs show strong evolution in luminosity.
 In the range of $0.48<z<2$, $L^*_{8\mu m}$ evolves as $\propto (1+z)^{1.6\pm0.2}$.
 Detailed comparison with the literature will be presented in $\S$\ref{discussion}.

\subsection{Bolometric IR luminosity density based on the 8$\mu$m LF}\label{sec:SFR8um}
Constraining the star formation history of galaxies as a function of redshift is a key to understanding galaxy formation in the Universe.
One of the primary purposes in computing IR LFs is to estimate the IR luminosity density, which in turn is a good estimator of the dust hidden cosmic star formation density \citep{1998ARA&A..36..189K}. Since dust obscuration is more important for more actively star forming galaxies at higher redshift, and such star formation cannot be observed in UV light, it is important to obtain IR-based estimate in order to fully understand the cosmic star formation history of the Universe.

 We estimate  the total infrared luminosity density by integrating the LF weighted by the luminosity.
First, we need to convert $L_{8\mu m}$ to the bolometric infrared luminosity.
The bolometric IR luminosity of a galaxy is produced by the thermal emission of its interstellar matter. In star-forming galaxies, the UV radiation produced by young stars heats the interstellar dust, and the reprocessed light is emitted in the IR. For this reason, in star-forming galaxies, the bolometric IR luminosity is a good estimator of the current SFR (star formation rate) of the galaxy.
\citet{2008A&A...479...83B} showed a strong correlation between $L_{8\mu m}$ and total infrared luminosity ($L_{TIR}$) for 372 local star-forming galaxies. 
The conversion given by \citet{2008A&A...479...83B} is:

\begin{eqnarray}\label{equation}
L_{TIR}=377.9\times (\nu L_\nu)^{0.83}_{rest 8\mu m} (\pm37\%)
\end{eqnarray}

\citet{2007ApJ...660...97C} further constrained the sample to luminous, high S/N galaxies ($\nu L_{\nu}^{8\mu m}>10^{10}L_{\odot}$ and S/N$>3$ in all MIPS bands) in order to better match their sample, and derived the following equation.

\begin{eqnarray}\label{8um_equation_caputi}
L_{TIR}=1.91\times (\nu L_\nu)^{1.06}_{rest 8\mu m} (\pm 55\%)
\end{eqnarray}

Since ours is also a sample of bright galaxies, we use this equation to convert  $L_{8\mu m}$ to $L_{TIR}$.
 Because the conversion is based on local star-forming galaxies, it is a concern if it holds at higher redshift or not. \citet{2008A&A...479...83B} checked this by stacking 24$\mu$m sources at $1.3<z<2.3$ in the GOODS fields to find the stacked sources are consistent with the local relation. They concluded that equation (\ref{equation}) is valid to link $L_{8\mu m}$ and $L_{TIR}$ at  $1.3<z<2.3$.  \citet{Takagi_PAH} also show that local $L_{7.7\mu m}$ vs $L_{TIR}$ relation holds true for IR galaxies at z$\sim$1 (see their Fig.10). \citet{2008ApJ...675.1171P} showed that $z\sim$2 sub-millimeter galaxies lie on the relation between $L_{TIR}$  and $L_{PAH,7.7}$ that has been established for local starburst galaxies. 
$S_{70}/S_{24}$ ratios of 70$\mu$m sources in \citet{2007ApJ...668...45P} are also consistent with local SED templates. These results suggest it is reasonable to use equation (\ref{8um_equation_caputi}) for our sample.

The conversion, however, has been the largest source of error in estimating  $L_{TIR}$ from  $L_{8\mu m}$. \citet{2008A&A...479...83B} themselves quote 37\% of uncertainty, and that \citet{2007ApJ...660...97C} report 55\% of dispersion around the relation. It should be kept in mind that the restframe $8\mu$m is sensitive to the star-formation activity, but at the same time, it is where the SED models have strongest discrepancies due to the complicated PAH emission lines. A detailed comparison of different conversions is presented in Fig.12 of  \citet{2007ApJ...660...97C}, who reported factor of $\sim$5 of differences among various models.

\noindent Then the 8$\mu$m LF is weighted by the $L_{TIR}$ and integrated to obtain TIR density.
For integration, we first fit an analytical function to the LFs.
In the literature, IR LFs were fit better by a double-power law \citep{2006MNRAS.370.1159B} or a double-exponential \citep{1990MNRAS.242..318S,2004ApJ...609..122P,2006A&A...448..525T,2005ApJ...632..169L} than a Schechter function, which declines too suddenlly at the high luminosity, underestimating
the number of bright galaxies.  
In this work, we fit the 8$\mu$m LFs using a double-power law \citep{2006MNRAS.370.1159B} as shown below.

\begin{equation}
 \label{eqn:lumfunc2p}
 \Phi(L)dL/L^{*} = \Phi^{*}\bigg(\frac {L}{L^{*}}\bigg)^{1-\alpha}dL/L^{*}, ~~~ (L<L^{*})
\end{equation}

\begin{equation}
 \label{eqn:lumfunc2p2}
 \Phi(L)dL/L^{*} = \Phi^{*}\bigg(\frac {L}{L^{*}}\bigg)^{1-\beta}dL/L^{*}, ~~~  (L>L^{*})
\end{equation}

\noindent First, the  double-power law is fitted to the lowest redshift LF at 0.38$<z<$0.58 to determine the normalization ($\Phi^{*}$) and slopes ($\alpha,\beta$). 
 For higher redshifts we do not have enough statistics to simultaneously fit 4 parameters ($\Phi^{*}$, $L^*$, $\alpha$, and $\beta$).  Therefore, we fixed the slopes and normalization at the local values and varied only $L^*$ at for the higher-redshift LFs.
 Fixing the faint-end slope is a common procedure with the depth of current IR satellite surveys \citep{2006MNRAS.370.1159B,2007ApJ...660...97C}.
 The stronger evolution in luminosity than in density found by previous work \citep{2005ApJ...630...82P,2005ApJ...632..169L} also justifies this parametrization. 
 Best fit parameters are presented in Table \ref{tab:fit_parameters}.
 Once the best-fit parameters are found, we integrate the double power law outside the luminosity range in which we have data to obtain estimate of the total infrared luminosity density, $\Omega_{TIR}$.

\begin{table*}
 \centering
 \begin{minipage}{180mm}
  \caption{Best fit parameters for 8,12$\mu$m and TIR LFs}\label{tab:fit_parameters}
  \begin{tabular}{@{}ccclllcccc@{}}
  \hline
   Redshift & $\lambda$ & $L^*$ ($L_{\odot}$)& $\Phi^*(\mathrm{Mpc^{-3} dex^{-1}})$ & $\alpha$ & $\beta$  \\ 
 \hline
 \hline
0.38$<$z$<$0.58  &   8$\mu$m & ($2.2^{+0.3}_{-0.1})\times 10^{10}$ & (2.1$^{+0.3}_{-0.4})\times 10^{-3}$ & 1.75$^{+0.01}_{-0.01}$   &   3.5$^{+0.2}_{-0.4}$ 	\\
0.65$<$z$<$0.90  &   8$\mu$m & ($2.8^{+0.1}_{-0.1} )\times 10^{10}$ & $2.1\times 10^{-3}$  &1.75 & 3.5	\\
1.1$<$z$<$1.4    &   8$\mu$m & ($3.3^{+0.2}_{-0.2} )\times 10^{10}$ & $2.1\times 10^{-3}$  &1.75 & 3.5	\\
1.8$<$z$<$2.2    &   8$\mu$m & ($8.2^{+1.2}_{-1.8} )\times 10^{10}$ & $2.1\times 10^{-3}$  &1.75 & 3.5	\\
 \hline
0.15$<$z$<$0.35  &   12$\mu$m & ($6.8^{+0.1}_{-0.1} )\times 10^{9}$ &  (4.2$^{+0.7}_{-0.6})\times 10^{-3}$ &1.20$^{+0.01}_{-0.02}$   &   2.9$^{+0.4}_{-0.2}$ \\
0.38$<$z$<$0.62  &   12$\mu$m & (11.7$^{+0.3}_{-0.5} )\times 10^{9}$ &  4.2$\times 10^{-3}$ &1.20   &   2.9 \\
0.84$<$z$<$1.16  &   12$\mu$m & (14$^{+2}_{-3} )\times 10^{9}$ &  4.2$\times 10^{-3}$ &1.20   &   2.9 \\
 \hline
0.2$<$z$<$0.5  &  Total  & (1.2$^{+0.1}_{-0.2} )\times 10^{11}$& (5.6$^{+1.5}_{-0.2})\times 10^{-4}$ &  1.8$^{+0.1}_{-0.4}$   &   3.0$^{+1.0}_{-1.0}$ 	\\
0.5$<$z$<$0.8  &  Total  & (2.4$^{+1.8}_{-1.6} )\times 10^{11}$ &5.6$\times 10^{-4}$ &  1.8   &   3.0	\\
0.8$<$z$<$1.2  &  Total  & (3.9$^{+2.3}_{-2.2} )\times 10^{11}$ &5.6$\times 10^{-4}$ &  1.8   &   3.0	\\
1.2$<$z$<$1.6  &  Total  & (14$^{+1}_{-2} )\times 10^{11}$      &5.6$\times 10^{-4}$ &  1.8   &   3.0	\\
 \hline
\end{tabular}
\end{minipage}
\end{table*}

\begin{figure}
\begin{center}
\includegraphics[scale=0.6]{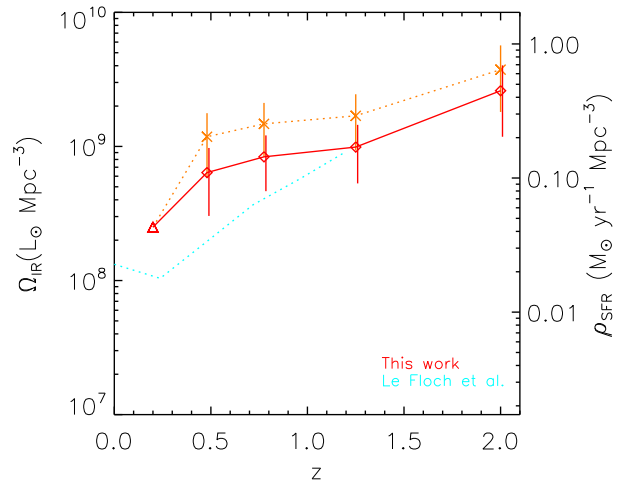}
\end{center}
\caption{
Evolution of TIR luminosity density computed by integrating the 8$\mu$m LFs in Fig.\ref{fig:8umlf}.The red solid lines use the conversion in equation (\ref{8um_equation_caputi}).
The orange dashed lines use equation (\ref{equation}).
Results from \citet{2005ApJ...632..169L} are shown with the cyan dotted lines.
}\label{fig:8umTLD}
\end{figure}

The resulting total luminosity density ($\Omega_{IR}$) is shown in Fig.\ref{fig:8umTLD} as a function of redshift. 
Errors are estimated by varying the fit within 1$\sigma$ of uncertainty in LFs, then errors in conversion from $L_{8\mu m}$ to $L_{TIR}$ are added. The latter is by far the larger source of uncertainty.
Simply switching from equation (\ref{equation}) (orange dashed line) to (\ref{8um_equation_caputi}) (red solid line) produces a $\sim$50\% difference.  
Cyan dashed lines show results from \citet{2005ApJ...632..169L} for a comparision. The lowest redshift point was corrected following \citet{2009A&A...496...57M}. 

\begin{figure}
\begin{center}
\includegraphics[scale=0.6]{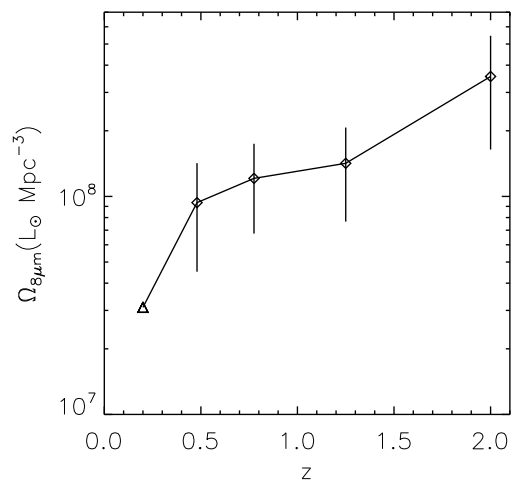}
\end{center}
\caption{
Evolution of 8$\mu$m IR luminosity density computed by integrating the 8$\mu$m LFs in Fig.\ref{fig:8umlf}. The lowest redshift point is from \citet{2007ApJ...664..840H}. 
}\label{fig:omega8um}
\end{figure}

We also show the evolution of monochromatic 8$\mu$m luminosity ($L_{8\mu m}$), which is obtained by integrating the fits, but without converting to $L_{TIR}$ in Fig.\ref{fig:omega8um}.
 The $\Omega_{8\mu m}$ evolves as $\propto (1+z)^{1.9\pm0.7}$.

 The SFR and $L_{TIR}$ are related by the following equation for a Salpeter IMF, 
$\phi$ (m) 
$\propto m^{-2.35}$ between 
$0.1-100 M_{\odot}$  \citep{1998ARA&A..36..189K}.
\begin{eqnarray}
SFR (M_{\odot} yr^{-1}) =1.72 \times 10^{-10} L_{TIR} (L_{\odot}) 
\end{eqnarray}

The right ticks of Fig.\ref{fig:8umTLD} shows the star formation density scale, converted from   $\Omega_{IR}$ using the above equation.

 In Fig.\ref{fig:8umTLD}, $\Omega_{IR}$ monotonically increases toward higher z.
 Compared with $z$=0, $\Omega_{IR}$ is $\sim$10 times larger at $z$=1.
The evolution between $z$=0.5 and $z$=1.2 is a little flatter, but this is perhaps due to a more irregular shape of LFs at 0.65$<z<$0.90, and thus, we do not consider it significant.
The results obtained here agree with previous work \citep[e.g.,][]{2005ApJ...632..169L} within the errors. We compare the results with previous work in more detail in $\S$\ref{discussion}.


\subsection{12$\mu$m LF}\label{sec:12umlf}

\begin{figure}
\begin{center}
\includegraphics[scale=0.6]{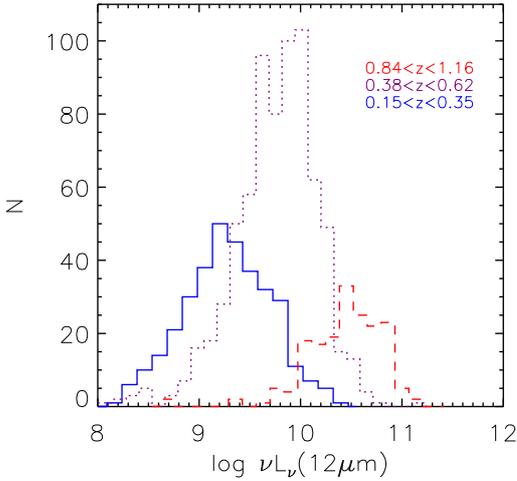}
\end{center}
\caption{
12$\mu$m luminosity distributions of samples used to compute  restframe  12$\mu$m LFs. 
From low redshift, 335, 573, and 213 galaxies are in each redshift bin.
}\label{fig:12um_luminosity}
\end{figure}

 In this subsection we estimate restframe 12$\mu$m LFs based on the AKARI NEP-Deep data.
 12$\mu$m luminosity ($L_{12\mu m}$) has been well studied through ISO and IRAS, and known to correlate closely with TIR luminosity \citep{1995ApJ...453..616S,2005ApJ...630...82P}. 
 
 As was the case for the 8$\mu$m LF, it is advantageous that AKARI's continuous filters in the mid-IR allow us to estimate restframe 12$\mu$m luminosity without much extrapolation based on SED models.
 
\begin{figure}
\begin{center}
\includegraphics[scale=0.6]{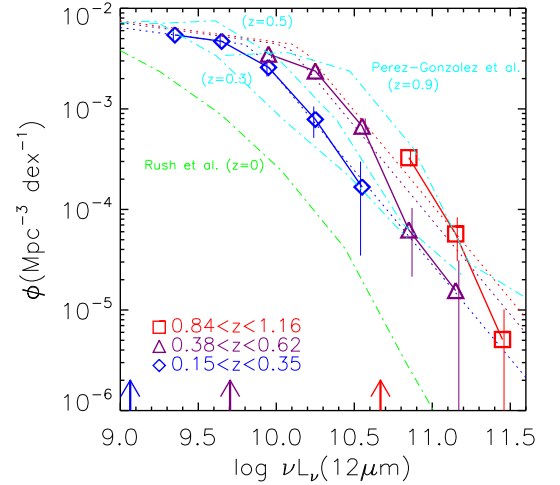}
\end{center}
\caption{
Restframe  12$\mu$m LFs based on the AKARI NEP-Deep field.
  The blue diamonds, purple triangles, and red squares show the 12$\mu$m LFs at $0.15<z<0.35, 0.38<z<0.62$, and $0.84<z<1.16$, respectively.
  Vertical arrows show the 12$\mu$m luminosity corresponding to the flux limit at the central redshift in each redshift bin.
  Overplotted are  \citet{2005ApJ...630...82P} at $z$=0.3,0.5 and 0.9 in the cyan dash-dotted lines, and \citet{1993ApJS...89....1R} at $z$=0 in the green dash-dotted lines.
AGNs are excluded from the sample ($\S$\ref{sec:photoz}).
}\label{fig:12umlf}
\end{figure}

 Targeted redshifts are $z$=0.25, 0.5 and 1 where $L15,L18W$ and $L24$ filters cover the restframe 12$\mu$m, respectively.
 We summarise the filters used in Table \ref{tab:filters_used}.
 Methodology is the same as for the 8$\mu$m LF; we used the sample to the 5$\sigma$ limit, corrected for the completeness, then used the 1/$V_{\max}$  method to compute LF in each redshift bin.
The histogram of $L_{12\mu m}$ distribution is presented in Fig.\ref{fig:12um_luminosity}.
 The resulting 12$\mu$m LF is shown in Fig.\ref{fig:12umlf}.
 Compared with \citet{1993ApJS...89....1R}'s $z$=0 LF based on IRAS Faint Source Catalog, the 12$\mu$m LFs show steady evolution with increasing redshift. 
 In the range of $0.25<z<1$, $L^*_{12\mu m}$ evolves as $\propto (1+z)^{1.5\pm0.4}$.

\subsection{Bolometric IR luminosity density based on the 12$\mu$m LF}\label{sec:SFR12um}

 12$\mu$m is one of the most frequently used monochromatic fluxes to estimate $L_{TIR}$.
 The total infrared luminosity is computed from the $L_{12\mu m}$ using the conversion in \citet{2001ApJ...556..562C,2005ApJ...630...82P}.

\begin{eqnarray}\label{eq:12um}
\log L_{TIR}=\log (0.89^{+0.38}_{-0.27})+1.094 \log L_{12\mu m}\label{Aug  7 17:32:13 2009}
\end{eqnarray}

\citet{2005A&A...432..423T} independently estimated the relation to be 

\begin{eqnarray}\label{eq:12um_takeuchi}
\log L_{TIR}=1.02+0.972 \log L_{12\mu m},
\end{eqnarray}

 which we also use to check our conversion. As both authors state, these conversions contain an error of factor of 2-3. Therefore, we should avoid conclusions that could be affected by such errors.

Then the 12$\mu$m LF is weighted by the $L_{TIR}$ and integrated to obtain TIR density.
Errors are estimated by varying the fit within 1$\sigma$ of uncertainty in LFs, and errors in converting from $L_{12\mu m}$ to $L_{TIR}$ are added. The latter is by far the largest source of uncertainty. 
 Best fit parameters are presented in Table \ref{tab:fit_parameters}.
In Fig.\ref{fig:12umTLD}, we show total luminosity density based on the 12$\mu$m LF presented in Fig.\ref{fig:12umlf}.
The results show a rapid increase of $\Omega_{IR}$, agreeing with previous work \citep{2005ApJ...632..169L} within the errors.

\begin{figure}
\begin{center}
\includegraphics[scale=0.6]{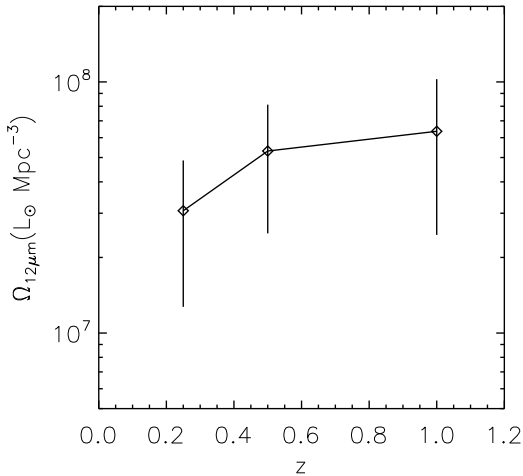}
\end{center}
\caption{
Evolution of 12$\mu$m IR luminosity density computed by integrating the 12$\mu$m LFs in Fig.\ref{fig:12umlf}.
}\label{fig:omega12um}
\end{figure}

\begin{figure}
\begin{center}
\includegraphics[scale=0.6]{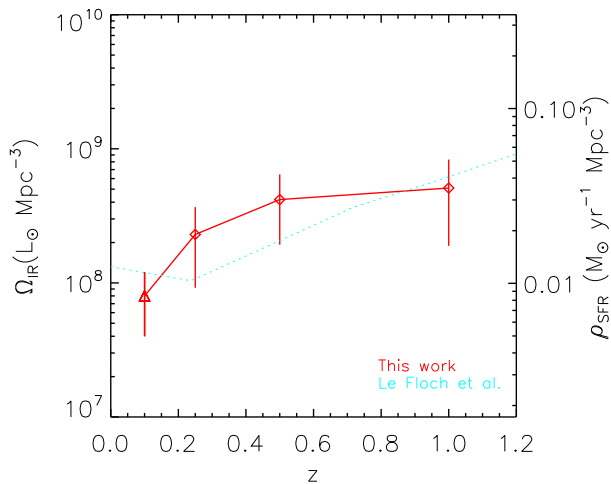}
\end{center}
\caption{
TIR luminosity density computed by integrating the 12$\mu$m LFs in Fig.\ref{fig:12umlf}.
}\label{fig:12umTLD}
\end{figure}

We also integrate monochromatic $L_{12\mu m}$ over the LFs (without converting to $L_{TIR}$) to derive the evolution of total $12\mu m$ monochromatic luminosity density, $\Omega_{12\mu m}$. The results are shown in Fig.\ref{fig:omega12um}, which shows a strong evolution of $\Omega_{12\mu m} \propto (1+z)^{1.4\pm1.4}$.
 It is interesting to compare this to $\Omega_{8\mu m} \propto (1+z)^{1.9\pm0.7}$ obtained in $\S$\ref{sec:SFR8um}.
 Although errors are significant on both estimates, $\Omega_{12\mu m}$ and $\Omega_{8\mu m}$ show a possibly different evolution, suggesting that the cosmic infrared spectrum changes its SED shape. Whether this is due to evolution in dust, or dusty AGN contribution is an interesting subject for future work.

\subsection{TIR LF} \label{sec:tirlf}

AKARI's continuous mid-IR coverage is also superior for SED-fitting to estimate $L_{TIR}$, since for star-forming galaxies, the mid-IR part of the IR SED is dominated by the PAH emissions which reflect the SFR of galaxies, and thus, correlates well with $L_{TIR}$, which is also a good indicator of the galaxy SFR. 
The AKARI's continuous MIR coverage helps us to estimate $L_{TIR}$.

\begin{figure}
\begin{center}
\includegraphics[scale=0.3]{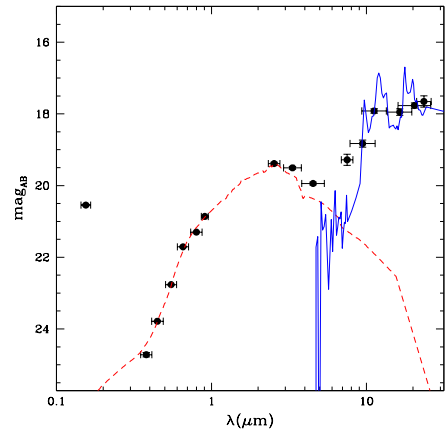}
\end{center}
\caption{An example of the SED fit. The red dashed line shows the best-fit SED for the UV-optical-NIR SED, mainly to estimate photometric redshift. The blue solid line shows the best-fit model for the IR SED at $\lambda>6\mu$m, to estimate $L_{TIR}$.
}\label{fig:SEDfit}
\end{figure}

\begin{figure}
\begin{center}
\includegraphics[scale=0.3]{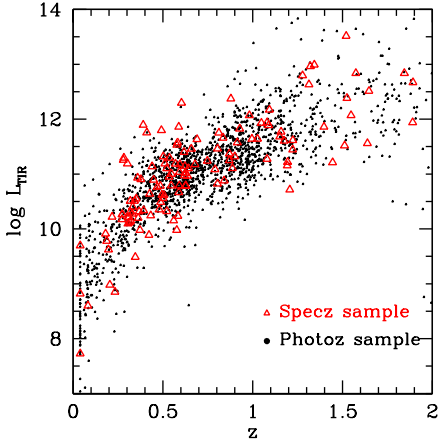}
\end{center}
\caption{
TIR luminosity is shown as a function of photometric redshift.
The photo-$z$ is estimated using UV-optical-NIR photometry.
$L_{TIR}$ is obtained through SED fit in 7-24$\mu$m.
}\label{fig:z_lir}
\end{figure}

After photometric redshifts are estimated using the UV-optical-NIR photometry, we fix the redshift at the photo-$z$, then use the same {\ttfamily  LePhare} code to fit the infrared part of the SED to estimate TIR luminosity. 
We used \citet{2003MNRAS.338..555L}'s SED templates to fit the photometry using the AKARI bands at $>$6$\mu$m ($S7,S9W,S11,L15,L18W$ and $L24$). 
We show an example of the SED fit in Fig. \ref{fig:SEDfit}, where the red dashed and blue solid lines show the best-fit SEDs for the UV-optical-NIR and IR SED at $\lambda>6\mu$m, respectively. 
 The obtained total infrared luminosity ($L_{TIR}$) is shown as a function of redshift in Fig.\ref{fig:z_lir}, with spectroscopic galaxies in large triangles. 
 The figure shows that the AKARI can detect LIRGs ($L_{TIR}>10^{11}L_{\odot}$) up to $z$=1 and ULIRGs ($L_{TIR}>10^{12}L_{\odot}$) to $z$=2.
 We also checked that using different SED models \citep{2001ApJ...556..562C,2002ApJ...576..159D}  does not change our essential results.

 \begin{figure}
 \begin{center}
 \includegraphics[scale=0.6]{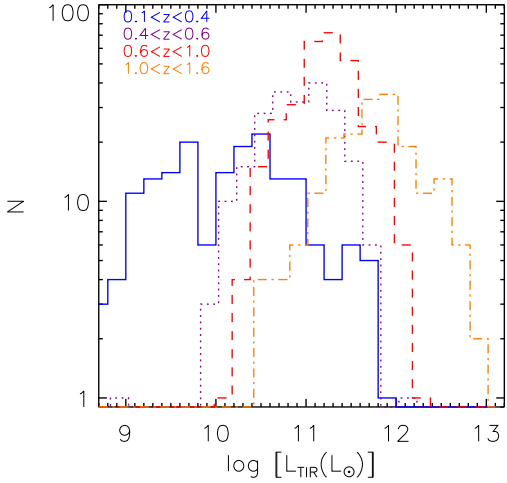}
 \end{center}
 \caption{
 A histogram of TIR luminosity.
 From low-redshift, 144,      192,      394, and      222 galaxies are in  0.2$<z<$0.5, 0.5$<z<$0.8, 0.8$<z<$1.2,  and 1.2$<z<$1.6, respectively.
 }\label{fig:tir_hist}
 \end{figure}

 Galaxies in the targeted redshift range are best sampled in the 18$\mu$m band due to the wide bandpass of the $L18W$ filter \citep{2006PASJ...58..673M}. In fact, in a single-band detection, the 18$\mu$m image returns the largest number of sources.
 Therefore, we applied the 1/$V_{\max}$  method using the detection limit at $L18W$.
 We also checked that using the $L15$ flux limit does not change our main results.
 The same \citet{2003MNRAS.338..555L}'s models are also used for $k$-corrections necessary to compute $V_{\max}$  and $V_{\min}$.
 The redshift bins used are 0.2$<z<$0.5, 0.5$<z<$0.8, 0.8$<z<$1.2,  and 1.2$<z<$1.6.  
  A distribution of $L_{TIR}$ in each redshift bin is shown in Fig.\ref{fig:tir_hist}.

 The obtained $L_{TIR}$ LFs are shown in Fig.\ref{fig:TIR_LF}.
 The uncertainties are esimated through the Monte Carlo simulations ($\S$\ref{sec:monte}).
 For a local benchmark, we overplot \citet{2003AJ....126.1607S} who derived LFs from the analytical fit to the IRAS Revised Bright Galaxy Sample, i.e., $\phi \propto L^{-0.6}$ for $L<L^*$ and $\phi \propto L^{-2.2}$ for $L>L^{*}$ with $L^*=10^{10.5}L_{\odot}$. 
 The TIR LFs show a strong evolution compared to local LFs. 
 At $0.25<z<1.3$, $L^*_{TIR}$ evolves as $\propto (1+z)^{4.1\pm0.4}$.
 We further compare LFs to the previous work in $\S$\ref{discussion}.
 
\subsection{Bolometric IR luminosity density based on the TIR LF}\label{sec:SFRTIR}

\begin{figure}
\begin{center}
\includegraphics[scale=0.6]{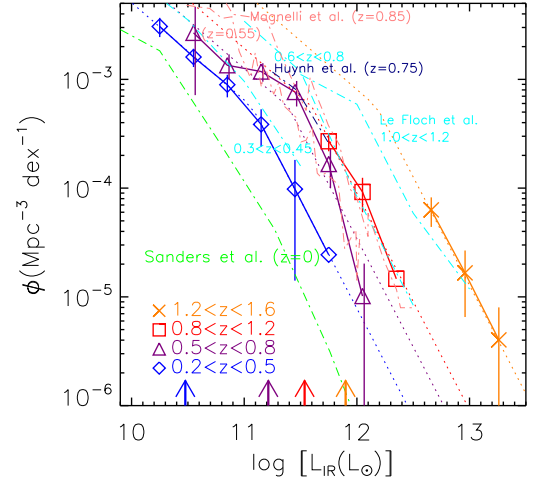}
\end{center}
\caption{
TIR LFs.
Vertical lines show the luminosity corresponding to the flux limit at the central redshift in each redshift bin.
AGNs are excluded from the sample ($\S$\ref{sec:photoz}).
}\label{fig:TIR_LF}
\end{figure}
\begin{figure}
\begin{center}
\includegraphics[scale=0.6]{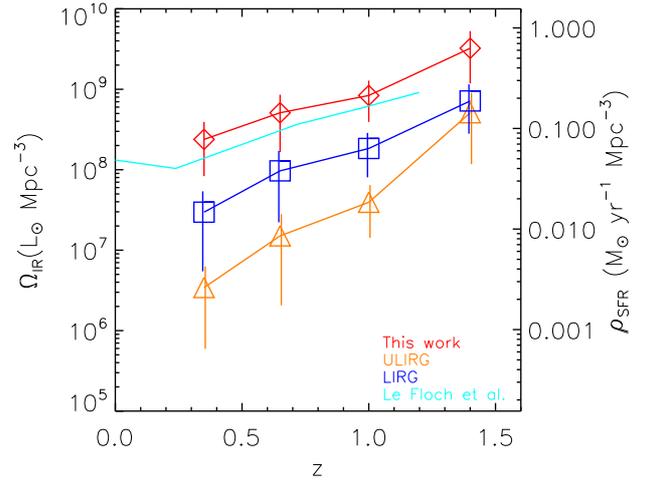}
\end{center}
\caption{
TIR luminosity density (red diamonds) computed by integrating the total LF in Fig.\ref{fig:TIR_LF}.
The blue squares and orange triangles are for LIRG and ULIRGs only.
}\label{fig:TLD}
\end{figure}

 Using the same methodology as in previous sections, 
 we integrate $L_{TIR}$ LFs in Fig.\ref{fig:TIR_LF} through a double-power law fit (eq. \ref{eqn:lumfunc2p} and \ref{eqn:lumfunc2p2}). The resulting evolution of the TIR density is shown with red diamonds in Fig.\ref{fig:TLD},
which in in good agreement with \citet{2005ApJ...632..169L} within the errors.
 Errors are estimated by varying the fit within 1$\sigma$ of uncertainty in LFs.
 For uncertainty in the SED fit, we added 0.15 dex of error.
 Best fit parameters are presented in Table \ref{tab:fit_parameters}.
 In Fig.\ref{fig:TLD}, we also show the contributions to $\Omega_{TIR}$ from LIRGs and ULIRGs with the blue squares and orange triangles, respectively.
 We further discuss the evolution of  $\Omega_{TIR}$ in $\S$\ref{discussion}.

\section{Discussion}
\label{discussion}

\subsection{Comparison with previous work}

 In this section, we compare our results to previous work, especially those based on the Spitzer data.  Comparisons are best done in the same wavelengths, since the conversion from either $L_{8\mu m}$ or  $L_{12\mu m}$ to $L_{TIR}$ involves the largest uncertainty.
 Hubble parameters in the previous work are converted to $h=0.7$ for comparison.

\subsubsection{8$\mu$m LFs}\label{sec:8umLF_comp}
 Recently, using the Spitzer space telescope, restframe 8$\mu$m LFs of $z\sim$1 galaxies have been computed in detail by \citet{2007ApJ...660...97C} in the GOODS fields and by \citet{2006MNRAS.370.1159B} in the SWIRE field. 
 In this section, we compare our restframe 8$\mu$m LFs (Fig.\ref{fig:8umlf}) to these and discuss possible differences.

 In Fig.\ref{fig:8umlf}, we overplot \citet{2007ApJ...660...97C}'s LFs at $z$=1 and $z$=2 in the cyan dash-dotted lines. Their $z$=2 LF is in good agreement with our LF at 1.8$<z<$2.2. 
 However, their $z$=1 LF is larger than ours by a factor of 3-5 at $\log L>11.2$. Note that the brightest ends ($\log L\sim 11.4$) are consistent with each other to within 1$\sigma$. They have excluded AGN using optical-to-X-ray flux ratio, and we also have excluded AGN through the optical SED fit. Therefore, especially at the faint-end, the contamination from AGN is not likely to be the main cause of differences. Since  \citet{2007ApJ...660...97C} uses GOODS fields, cosmic variance may play a role here.
 The exact reason for the difference is unknown, but we point out that their $\Omega_{IR}$ estimate at $z$=1 is also higher than other estimates by a factor of a few (see their Fig.15).
 Once converted into $L_{TIR}$, \citet{2009A&A...496...57M} also reported \citet{2007ApJ...660...97C}'s  $z$=1 LF is higher than their estimate based on 70$\mu$m by a factor of several (see their Fig.12). They concluded the difference is from different SED models used, since their LF matched with that of \citet{2007ApJ...660...97C}'s once the same SED models were used. 
 We will compare our total LFs to those in the literature below.
 
  \citet{2006MNRAS.370.1159B} also computed restframe 8$\mu$m LFs using the Spitzer/SWIRE data. We overplot their results at $0.25<z<0.5$ and $0.5<z<1$ in Fig.\ref{fig:8umlf} with the pink dot-dashed lines. 
 In both redshift ranges, good agreement is found at higher luminosity bins ($L_{8\mu m}>10^{10.5} L_{\odot}$). 
 However, at all redshift ranges including the ones not shown here, \citet{2006MNRAS.370.1159B} tends to show a flatter faint-end tail than ours, and a smaller $\phi$ by a factor of $\sim$3. Although the exact reason is unknown, the deviation starts toward the fainter end, where both works approach the flux limits of the surveys. Therefore, possibly incomplete sampling may be one of the reasons. 
 It is also reported that the faint-end of IR LFs depends on the environment, in the sense that higher-density environment has steeper faint-end tail \citep{cluster_LF}.
 Note that at $z$=1, \citet{2006MNRAS.370.1159B}'s LF (pink) deviates from that by \citet{2007ApJ...660...97C} (cyan) by almost a magnitude. Our  8$\mu$m LFs are between these works.

These comparisons suggest that even with the current generation of satellites and state-of-the-art SED models, factor-of-several uncertainties still remain in estimating the 8$\mu$m LFs at z$\sim$1. 
More accurate determination has to await a larger and deeper survey by the next generation IR satellites such as Herschel and WISE.

 To summarise,  our 8$\mu$m LFs are between those by \citet{2006MNRAS.370.1159B} and \citet{2007ApJ...660...97C}, and discrepancy is by a factor of several at most. 
 We note that both of the previous works had to rely on SED models to estimate $L_{8\mu m}$ from the Spitzer $S_{24\mu m}$ in the MIR wavelengths where SED modeling is difficult. Here, AKARI's mid-IR bands are advantageous in  directly observing redshifted restframe 8$\mu$m flux in one of the AKARI's filters, leading to more reliable measurement of 8$\mu$m LFs without uncertainty from the SED modeling.

\subsubsection{12$\mu$m LFs}

 \citet{2005ApJ...630...82P} investigated the evolution of restframe 12$\mu$m LFs using the Spitzer CDF-S and HDF-N data.
 We overplot their results in similar redshift ranges as the cyan dot-dashed lines in Fig.\ref{fig:12umlf}.
 Considering both LFs have significant error bars, these LFs are in good agreement with our LFs, and show significant evolution in the 12$\mu$m LFs compared with the $z$=0 12$\mu$m LF by \citet{1993ApJS...89....1R}. 
 The agreement is in a stark contrast to the comparison in 8$\mu$m LFs in $\S$\ref{sec:8umLF_comp}, where we suffered from differnces of a factor of several. A possible reason for this is that 12$\mu$m is sufficiently redder than 8$\mu$m, that it is easier to be extrapolated from $S_{24\mu m}$ in case of the Spitzer work. In fact, at $z$=1, both the Spitzer 24$\mu$m band and AKARI $L24$ observe the restframe 12$\mu$m directly. In additon, mid-IR SEDs around 12$\mu$m are flatter than at 8$\mu$m, where PAH emissions are prominent. Therefore, SED models can  predict the flux more accurately. In fact, this is part of the reason why  \citet{2005ApJ...630...82P} chose to investigate 12$\mu$m LFs. \citet{2005ApJ...630...82P} used \citet{2001ApJ...556..562C}'s SED to extrapolate $S_{24\mu m}$, and yet, they agree well with AKARI results, which are derived from filters that cover the restframe 12$\mu$m.
However, in other words, the discrepancy in 8$\mu$m LFs highlights the fact that the SED models are perhaps still imperfect in the 8$\mu$m wavelength range, and thus, MIR-spectroscopic data that covers wider luminosity and redshift ranges will be necessary to refine SED models in the mid-IR. AKARI's mid-IR slitless spectroscopy survey \citep{2008cosp...37.3370W} may help in this regard.

\subsubsection{TIR LFs}

 Lastly, we compare our TIR LFs (Fig.\ref{fig:TIR_LF}) with those in the literature. 
 Although the TIR LFs can also be obtained by converting 8$\mu$m LFs or 12$\mu$m LFs, we already compared our results in these wavelengths in the last subsections. Here, we compare our TIR LFs to \citet{2005ApJ...632..169L} and \citet{2009A&A...496...57M}.

\citet{2005ApJ...632..169L} obtained TIR LFs using the Spitzer CDF-S data. They have used the best-fit SED among various templates to estimate $L_{TIR}$. We overplot their total LFs in Fig.\ref{fig:TIR_LF} with the cyan dash-dotted lines. Only LFs that overlap with our redshit ranges are shown. The agreement at $0.3<z<0.45$ and $0.6<z<0.8$ is reasonable, considering the error bars on both sides. However, in all three redshift ranges, \citet{2005ApJ...632..169L}'s LFs are higher than ours, especially for $1.0<z<1.2$. 

 We also overplot TIR LFs by \citet{2009A&A...496...57M}, who used Spitzer 70$\mu$m flux and \citet{2001ApJ...556..562C}'s model to estimate $L_{TIR}$. In the two bins (centered on $z$=0.55 and $z$=0.85; pink dash-dotted lines) which closely overlap with our redshift bins, excellent agreement is found. 
 We also plot \citet{2007ApJ...667L...9H}'s LF at $0.6<z<0.9$ in the navy dash-dotted lines, which is computed from Spitzer $70\mu$m imaging in the GOODS-N, and this also shows very good agreement with ours.  
These LFs are on top of each other within the error bars, despite the fact that these measurements are from different data sets using different analyses.

This means that \citet{2005ApJ...632..169L}'s LFs is also higher than that of \citet{2009A&A...496...57M}, in addition to ours. A possible reason is that both \citet{2009A&A...496...57M} and we removed AGN (at least bright ones), whereas \citet{2005ApJ...632..169L} included them. This also is consistent with the fact that the difference is larger at $1.0<z<1.2$ where both surveys are only sensitive to luminous IR galaxies, which are dominated by AGN.  Another possible source of uncertainty is that \citet{2009A&A...496...57M} and we used a single SED library, while \citet{2005ApJ...632..169L} picked the best SED template among several libraries for each galaxy.

\subsection{Evolution of $\Omega_{IR}$}

\begin{figure*}
\begin{center}
\includegraphics[scale=0.8]{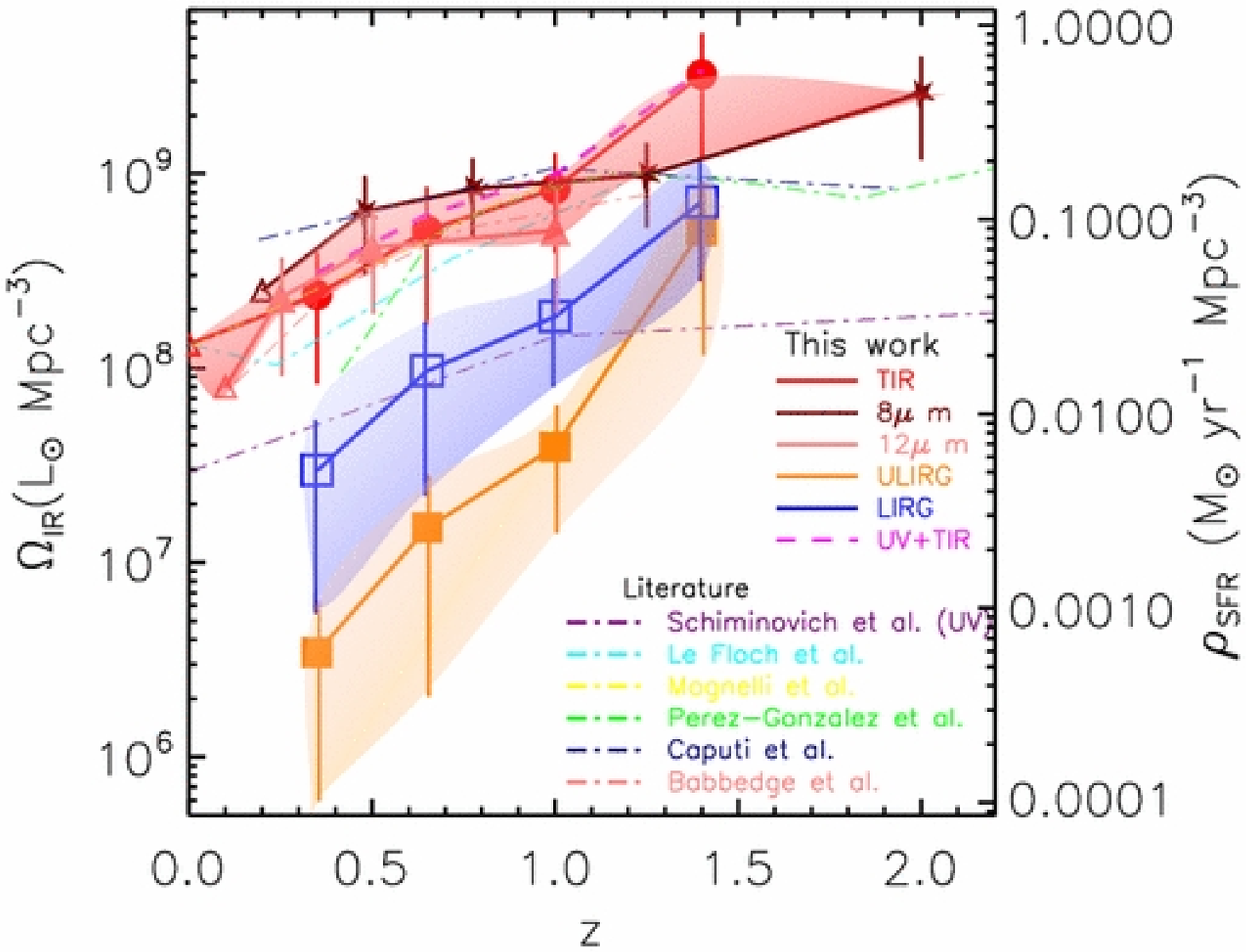}
\end{center}
\caption{
Evolution of TIR luminosity density based on TIR LFs (red circles), 8$\mu$m LFs (stars), and 12$\mu$m LFs (filled triangles). The blue open squares and orange filled squares  are for LIRG and ULIRGs only, also based on our $L_{TIR}$ LFs.
Overplotted dot-dashed lines are estimates from the literature: \citet{2005ApJ...632..169L}, \citet{2009A&A...496...57M} , \citet{2005ApJ...630...82P}, \citet{2007ApJ...660...97C},   and \citet{2006MNRAS.370.1159B} are in cyan, yellow, green, navy, and pink, respectively.
The purple dash-dotted line shows UV estimate by \citet{2005ApJ...619L..47S}.
The pink dashed line shows the total estimate of IR (TIR LF) and UV \citep{2005ApJ...619L..47S}.
}\label{fig:TLD_all}
\end{figure*}

In this section, we compare the evolution of $\Omega_{IR}$ as a function of redshift. In Fig.\ref{fig:TLD_all}, we plot $\Omega_{IR}$ estimated from  TIR LFs (red circles), 8$\mu$m LFs (brown stars), and 12$\mu$m LFs (pink filled triangles), as a function of redshift. Estimates based on 12$\mu$m LFs  and TIR LFs agree each other very well, while those from 8$\mu$m LFs show a slightly higher value by a factor of a few than others. This perhaps reflects the fact that 8$\mu$m is a more difficult part of the SED to be modeled, as we had a poorer agreement among papers in the literature in 8$\mu$m LFs . The bright-end slope of the double-power law was $3.5^{+0.2}_{-0.4}$ in Table \ref{tab:fit_parameters}. This is flatter than a Schechter fit by \citet{2006MNRAS.370.1159B} and a double-exponential fit by  \citet{2007ApJ...660...97C}. This is perhaps why we obtained higher $\Omega_{IR}$ in 8$\mu$m.

We overplot estimates from various papers in the literature \citep{2005ApJ...632..169L,2006MNRAS.370.1159B,2007ApJ...660...97C,2005ApJ...630...82P,2009A&A...496...57M} in the dash-dotted lines. Our $\Omega_{IR}$ has very good agreement with these at $0<z<1.2$, with almost all the dash-dotted lines lying within our error bars of  $\Omega_{IR}$ from $L_{TIR}$ and 12$\mu$m LFs. This is perhaps because an estimate of an integrated value such as $\Omega_{IR}$ is more reliable than that of LFs. 
 
 At $z>1.2$, our $\Omega_{IR}$ shows a hint of continuous increase, while \citet{2007ApJ...660...97C} and \citet{2006MNRAS.370.1159B} observed a slight decline at $z>1$. However, as both authors also pointed out, at this high-redshift range, both the AKARI and Spitzer satellites are sensitive to only LIRGs and ULIRGs, and thus the  extrapolation to fainter luminosities assumes the faint-end slope of the LFs, which could be uncertain. 
 In addition, this work has a poorer photo-z estimate at $z>0.8$ ($\frac{\Delta z}{1+z}$=0.10) due to the relatively shallow near-IR data.
Several authors tried to overcome this problem by stacking undetected sources. However, if an undetected source is also not detected at shorter wavelengths where positions for stacking are obtained,  it would not be included in the stacking either. Next generation satellite such as Herschel, WISE, and SPICA \citep{2008SPIE.7010E..15N} will determine the faint-end slope at $z>1$ more precisely.

We parameterize the evolution of $\Omega_{IR}$ using a following function.

\begin{equation} \label{eqn:SFRevolution}
\Omega_{IR}(z) \propto (1+z)^{\gamma}
\end{equation}

\noindent By fitting this to the $\Omega_{IR}$ from TIR LFs, we obtained 
$\gamma = 4.4\pm 1.0$.  This is consistent with most previous works. For example, 
\citet{2005ApJ...632..169L} obtained $\gamma = 3.9\pm 0.4$,
\citet{2005ApJ...630...82P} obtained $\gamma = 4.0\pm 0.2$,
\citet{2006MNRAS.370.1159B} obtained $\gamma = 4.5^{+0.7}_{-0.6}$,
\citet{2009A&A...496...57M} obtained $\gamma = 3.6\pm 0.4$.
The agreement was expected from Fig.\ref{fig:TLD_all}, but confirms a strong evolution of $\Omega_{IR}$. 

%


\subsection{Differential evolution among ULIRG, LIRG, normal galaxies}

In Fig. \ref{fig:TLD}, we also plot the contributions to $\Omega_{IR}$ from LIRGs and ULIRGs (measured from TIR LFs) with the blue open squares and orange filled squares, respectively. Both LIRGs and ULIRGs show strong evolution, as has been seen for total $\Omega_{IR}$ in the red filled circles. 
 Normal galaxies ($L_{TIR}<10^{11} L_{\odot}$) are still dominant, but decrease their contribution toward higher redshifts. In contrast, ULIRGs continue to increase their contribution.
 From $z$=0.35 to $z$=1.4, $\Omega_{IR}$ by LIRGs increases by a factor of $\sim$1.6, and 
  $\Omega_{IR}$ by ULIRGs increases by a factor of $\sim$10.
 The physical origin of ULIRGs in the local Universe is often merger/interaction  \citep{1996ARA&A..34..749S,1998ApJ...501L.167T,2005MNRAS.360..322G}. It would be interesting to investigate 
 whether the merger rate also increases in proportion to the ULIRG fraction, or if different mechanisms can also produce ULIRGs at higher redshift.

\subsection{Comparison to the UV estimate}

\begin{figure}
\begin{center}
\includegraphics[scale=0.6]{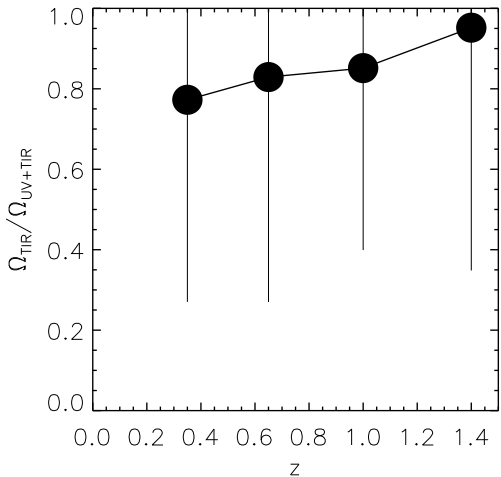}
\end{center}
\caption{
Contribution of $\Omega_{TIR}$ to $\Omega_{total}=\Omega_{UV}+\Omega_{TIR}$ is shown as a function of redshift.
}\label{fig:UVTIR}
\end{figure}


We have been emphasizing the importance of IR probes of the total SFRD of the Universe. 
However, the IR estimates do not take into account the contribution of the unabsorbed UV light produced by the young stars. Therefore, it is important to estimate how significant this UV contribution is.
 
\citet{2005ApJ...619L..47S} found that the energy density measured at 1500\AA~ evolves as $\propto (1+z)^{2.5\pm0.7}$ at $0<z<1$ and $\propto (1+z)^{0.5\pm0.4}$ at $z>1$.
using the GALEX data supplemented by the VVDS spectroscopic redshifts. We overplot their UV estimate of $\rho_{SFR}$ with the purple dot-dashed line in Fig.\ref{fig:TLD_all}. The UV estimate is almost a factor of 10 smaller than the IR estimate at most of the redshifts, confirming the importance of IR probes when investing the evolution of the total cosmic star formation density. 
In Fig.\ref{fig:TLD_all} we also plot total SFD (or $\Omega_{total}$) by adding $\Omega_{UV}$ and $\Omega_{TIR}$, with the magenta dashed line. 
 In Fig.\ref{fig:UVTIR}, we show the ratio of the IR contribution to the total SFRD of the Universe ($\Omega_{TIR}$/ $\Omega_{TIR}+\Omega_{UV}$) as a function of redshift. Although the errors are large,  Fig.\ref{fig:UVTIR} agrees with \citet{2005A&A...440L..17T}, and suggests that $\Omega_{TIR}$ explains 70\% of $\Omega_{total}$ at $z$=0.25, and that by $z$=1.3, 90\% of the cosmic SFD is explained by the infrared. This implies that $\Omega_{TIR}$ provides good approximation of the  $\Omega_{total}$ at $z>1$.

\section{Summary}

 We have estimated restframe 8$\mu$m, 12$\mu$m, and total infrared luminosity functions using the AKARI NEP-Deep data.
 Our advantage over previous work is AKARI's continuous filter coverage in the mid-IR wavelengths (2.4, 3.2, 4.1, 7, 9, 11, 15, 18, and 24$\mu$m), which allow us to estimate mid-IR luminosity without a large extrapolation based on SED models, which were the largest uncertainty in previous work. Even for $L_{TIR}$, the SED fitting is much more reliable due to this continuous coverage of mid-IR filters. 

 Our findings are as follows:
\begin{itemize}
 \item 8$\mu$m LFs show a strong and continuous evolution from $z$=0.35 to $z$=2.2. Our LFs are larger than \citet{2006MNRAS.370.1159B}, but smaller than \citet{2007ApJ...660...97C}. The difference perhaps stems from the different SED models, highlighting a difficulty in SED modeling at wavelengths crowded by strong PAH emissions.
 $L^*_{8\mu m}$ shows a continuous evolution as $L^*_{8\mu m}$ $\propto (1+z)^{1.6\pm0.2}$ in the range of $0.48<z<2$.
 \item 12$\mu$m LFs show a strong and continuous evolution from $z$=0.15 to $z$=1.16 with $L^*_{12\mu m}$ $\propto (1+z)^{1.5\pm0.4}$. This agrees well with \citet{2005ApJ...630...82P}, including a flatter faint-end slope. A better agreement than with 8$\mu$m LFs was obtained, perhaps because of smaller uncertainty in modeling  the 12$\mu$m SED, and less extrapolation  needed in Spitzer 24$\mu$m observations.
 \item The TIR LFs show good agreement with \citet{2009A&A...496...57M}, but are smaller than \citet{2005ApJ...632..169L}.  At $0.25<z<1.3$, $L^*_{TIR}$ evolves as $\propto (1+z)^{4.1\pm0.4}$.
 Possible causes of the disagreement include different treatment of SED models in estimating $L_{TIR}$, and AGN contamination.
 \item TIR densities estimated from 12$\mu$m and TIR LFs show a strong evolution as a function of redshift, with $\Omega_{IR}\propto (1+z)^{4.4\pm 1.0}$. $\Omega_{IR}(z)$ also show an excellent agreement with previous work at $z<1.2$.
 \item We investigated the differential contribution to $\Omega_{IR}$ by ULIRGs and LIRGs.
 We found that the ULIRG (LIRG) contribution increases by a factor of 10 (1.8) from $z$=0.35 to $z$=1.4, suggesting IR galaxies are more dominant source of $\Omega_{IR}$ at higher redshift.
 \item We estimated that  $\Omega_{IR}$ captures 80\% 
 of the cosmic star formation at redshifts less than 1, and virtually all of it at higher redshift.  
 Thus adding the unobscured star formation detected at UV wavelengths would not change SFRD estimates significantly.
\end{itemize}

\section*{Acknowledgments}

We are grateful to  S.Arnouts for providing the LePhare code, and kindly helping us in using the code.
We thank the anonymous referee for many insightful comments, which significantly improved the paper.


T.G. and H.I. acknowledge financial support from the Japan Society for the Promotion of Science (JSPS) through JSPS Research Fellowships for Young Scientists.
HML acknowledges the support from KASI through its cooperative fund in 2008. 
TTT has been supported by Program for Improvement of Research
Environment for Young Researchers from Special Coordination Funds for
Promoting Science and Technology, and the Grant-in-Aid for the Scientific
Research Fund (20740105) commissioned by the Ministry of Education,
Culture,
Sports, Science and Technology (MEXT) of Japan.
TTT has been also partially supported from the Grand-in-Aid for the Global
COE Program ``Quest for Fundamental Principles in the Universe: from
Particles to the Solar System and the Cosmos'' from the MEXT.



%

This research is based on the observations with AKARI, a JAXA project with the participation of ESA.

The authors wish to recognize and acknowledge the very significant cultural role and reverence that the summit of Mauna Kea has always had within the indigenous Hawaiian community.  We are most fortunate to have the opportunity to conduct observations from this sacred mountain.
%
%





\label{lastpage}

\end{document}